\documentclass[aps,prd,tightenlines,preprint,nofootinbib,a4paper]{revtex4}
\usepackage{epsfig}

\begin{document}

\title{Rephasing invariance and the neutrino $\mu-\tau$ symmetry}

\author{S. H. Chiu\footnote{schiu@mail.cgu.edu.tw}}
\affiliation{Physics Group, CGE, Chang Gung University, 
Kwei-Shan 333, Taiwan}

\author{T. K. Kuo\footnote{tkkuo@purdue.edu}}
\affiliation{Department of Physics, Purdue University, West Lafayette, IN 47907, USA}

\begin{abstract}

The vacuum neutrino mixing is known to exhibit an approximate 
$\mu-\tau$ symmetry, which was shown to be preserved for
neutrino propagating in matter.  This symmetry reduces the neutrino
transition probabilities to very simple forms when expressed
in a rephasing invariant parametrization introduced earlier.
Applications to long baseline experiments are discussed.

\end{abstract}


\maketitle

\pagenumbering{arabic}



\section{Introduction}

The tremendous progress in the last decade has made it possible to pin down,
with impressive accuracy, many of the fundamental parameters in
the neutrino sector.  A complete picture, however, is still not available.
Chief among the missing information is the determination of the
$V_{13}$ element of the neutrino mixing matrix $V$, which, in turn,
is crucial in ascertaining the CP violation effects in the leptonic sector.
Given that direct CP violations in the quark sector \cite{had} have been well-established  
and accurately measured, it is imperative, from both the theoretical and experimental
points of view, to assess the corresponding situation in the leptonic sector.
Another unsolved puzzle concerns the neutrino mass spectrum, in that there 
are the possibilities of either the ``normal" or ``inverted" orderings.
It is certainly important to settle this question.

While the fundamental parameters refer to those in vacuum,
it has been well-established (see, \emph{e.g.}, Ref.\cite{wolf,ms,matter2,matter3,matter4,
matter5,matter6,matter7,matter8,matter9,Ki,cpm11,cpm12,cpm13}) 
that they are modified when neutrinos propagate through matter,
by giving the  neutrino an induced mass, which is proportional to its
energy and to the medium density.  Indeed, in the analyses of the solar neutrinos, 
certain features of the data, such as the modification of the energy spectra from the original, 
can only be understood by the inclusion of matter effects.  With the advent of long
baseline experiments (LBL, for an incomplete list, see, \emph{e.g.},
Ref.\cite{LBL,BMW,CHA,gold1,gold2,AK,silver1,silver2}), the induced mass can actually 
be ``tuned" by changing the neutrino
energy ($E$).  This provides a powerful tool which can be used to extract fundamental
neutrino parameters from measurements.

In this work, we will use a rephasing invariant parametrization which enables 
us to obtain simple formulas for the transition probabilities of neutrinos propagating 
through matter of constant density.  It was shown earlier that these parameters obey evolution
equations as a function of the induced mass.  In addition, these equations preserve the
approximate $\mu-\tau$ symmetry \cite{23-sym,LAM} which characterizes the neutrino mixing in vacuum.
Incorporation of the $\mu-\tau$ symmetry for all induced mass values results in a set
of very simple transition probabilities $P(\nu_{\alpha} \rightarrow \nu_{\beta})$.
In general, these formulas offer quick estimates of the various oscillation
probabilities, using the known solutions obtained earlier.
As an example, we will analyze $P(\nu_{e}\rightarrow \nu_{\mu})$ in detail,
emphasizing its dependence on the neutrino parameters.



\section{The rephrasing invariant parametrization}

Neutrino oscillations, being lepton-number conserving, are described in
terms of a mixing matrix whose possible Majorana phases are not observable.
Thus it behaves just like the CKM matrix under rephasing transformations,
which leave physical observables invariant \cite{Kuo:05}. 
To date, however, such observables
are often given in terms of parameters which are not individually invariant.
So it seems that the use of manifestly invariant parameters
may be more physically relevant.  Two such sets are known to be
$|V_{ij}|$ \cite{CH,Branco} and $V_{\alpha i}V_{\beta j}V^{*}_{\alpha j}V^{*}_{\beta i}$ \cite{Jar:85}.
Recently, by imposing the condition $\mbox{det}V= +1$ (without loss
of generality), another set was found, given by \cite{Kuo:05,Chiu:09,CKL,CK10} 
\begin{equation}\label{eq:g}
\Gamma_{ijk}=V_{1i}V_{2j}V_{3k}=R_{ijk}-iJ,
\end{equation}
where the common imaginary part can be identified with
the Jarlskog invariant $J$ \cite{Jar:85}.  
Their real parts are labeled as
\begin{equation}
(R_{123},R_{231},R_{312};R_{132},R_{213},R_{321})
=(x_{1},x_{2},x_{3};y_{1},y_{2},y_{3}).
\end{equation}
The variables are bounded by 
$-1 \leq (x_{i},y_{j}) \leq +1$
with $y_{j} \leq x_{i}$ for any ($i,j$), and satisfy two constraints:
\begin{equation}\label{cons}
\mbox{det}V=(x_{1}+x_{2}+x_{3})-(y_{1}+y_{2}+y_{3})=1,
\end{equation}
\begin{equation}\label{con2}
(x_{1}x_{2}+x_{2}x_{3}+x_{3}x_{1})-(y_{1}y_{2}+y_{2}y_{3}+y_{3}y_{1})=0.
\end{equation}
Eq.~(\ref{con2}), together with the relation
\begin{equation}\label{eq:J}
J^{2}=x_{1}x_{2}x_{3}-y_{1}y_{2}y_{3},
\end{equation}
follow \cite{Kuo:05} from (the imaginary and real parts of) the identity 
$\Gamma_{123}\Gamma_{231}\Gamma_{312}=\Gamma_{132}\Gamma_{213}\Gamma_{321}$.
Thus, flavor mixing is specified by the set $(x,y)$ plus a sign,
according to $J=\pm\sqrt{J^{2}}$.  This sign arises since the
transformation $V \rightarrow V^{*}$, corresponding to a CP conjugation,
leaves the real part $(x,y)$ of $\Gamma_{ijk}$ invariant, but changes
the sign of its imaginary part $(J)$.  Note that, using $|V_{ij}|^{2}$,
a complete parametrization also requires four $|V_{ij}|^{2}$ elements plus a sign.


The parameters $(x,y)$ are related to the rephasing invariant elements $|V_{ij}|^{2}$ by
\begin{equation}\label{eq:w}
 W = \left[|V_{ij}|^{2}\right]
   = \left(\begin{array}{ccc}
                    x_{1}-y_{1} & x_{2}-y_{2}   &  x_{3}-y_{3} \\
                     x_{3}-y_{2} & x_{1}-y_{3}  & x_{2}-y_{1} \\
                    x_{2}-y_{3}  &   x_{3}-y_{1}    & x_{1}-y_{2} \\
                    \end{array}\right).  
\end{equation}
One can readily obtain the parameters $(x,y)$ from $W$ by computing its cofactors,
which form the matrix $w$ with $w^{T}W=(\mbox{det}W)I$, and is given by
\begin{equation}\label{eq:co}
 w = \left(\begin{array}{ccc}
                    x_{1}+y_{1} & x_{2}+y_{2}   &  x_{3}+y_{3} \\
                     x_{3}+y_{2} & x_{1}+y_{3}  & x_{2}+y_{1} \\
                    x_{2}+y_{3}  &   x_{3}+y_{1}    & x_{1}+y_{2} \\
                    \end{array}\right).     
\end{equation}

The relations between $(x,y)$ and 
\begin{equation}\label{Pi}
\Pi^{\alpha \beta}_{ij} \equiv V_{\alpha i}V_{\beta j}V^{*}_{\alpha j}V^{*}_{\beta i}
\end{equation}
are given by (using $V_{\alpha i}V_{\beta j}-V_{\alpha j}V_{\beta i}
=\sum_{\gamma k} \epsilon_{\alpha \beta \gamma}\epsilon_{ijk}V^{*}_{\gamma k}$):
\begin{eqnarray}\label{Pi2}
\Pi^{\alpha \beta}_{ij} &=& |V_{\alpha i}|^{2}|V_{\beta j}|^{2}-
\sum_{\gamma k} \epsilon_{\alpha \beta \gamma}\epsilon_{ijk}V_{\alpha i}V_{\beta j}V_{\gamma k} \nonumber \\
&=& |V_{\alpha j}|^{2}|V_{\beta i}|^{2}+
\sum_{\gamma k} \epsilon_{\alpha \beta \gamma}\epsilon_{ijk}V^{*}_{\alpha j}V^{*}_{\beta i}V^{*}_{\gamma k}.
\end{eqnarray}
The second term in either expression is one of the $\Gamma$'s ($\Gamma^{*}$'s) defined in 
Eq.~(\ref{eq:g}).  Also, by using the constraint in Eq.~(\ref{cons}),
$\mbox{Re}(\Pi^{\alpha \beta}_{ij})$ can be expressed in terms of quadratics in $(x,y)$,
a result which will be used later in Tables I and II.



\section{Evolution equations and the $\mu-\tau$ symmetry}

For neutrinos in matter (of constant density),
it was shown \cite{CKL,CK10} that, as a function of the induced mass   
$A=2\sqrt{2}G_{F}n_{e}E$, the neutrino parameters satisfy a set of
evolution equations which are greatly simplified by using the $(x,y)$ variables.
It was found that 
\begin{equation}\label{eq:di}
\frac{dD_{i}}{dA}= |V_{1i}|^{2}=x_{i}-y_{i}, \hspace{.2in} (i=1,2,3)
\end{equation}  
where $D_{i}$ are the eigenvalues of the Hamiltonian.
Also, the evolution equations for all $(x_{i},y_{j})$ can be obtained
and are collected in Table I of Ref. \cite{CKL,CK10}.   
Of particular interest for our purposes are the equations:
\begin{eqnarray}
\frac{d\ln J}{dA}&=&\frac{-(x_{1}-y_{1})+(x_{2}-y_{2})}{D_{1}-D_{2}}+
\frac{-(x_{2}-y_{2})+(x_{3}-y_{3})}{D_{2}-D_{3}} \nonumber \\
&+&\frac{(x_{1}-y_{1})-(x_{3}-y_{3})}
{D_{3}-D_{1}},
\end{eqnarray}
and
\begin{eqnarray}\label{eq:ln}
\frac{1}{2}\frac{d}{dA}\ln(x_{1}-y_{1})& = & \frac{x_{2}-y_{2}}{D_{1}-D_{2}}-
\frac{x_{3}-y_{3}}{D_{3}-D_{1}}, \nonumber \\
\frac{1}{2}\frac{d}{dA}\ln(x_{2}-y_{2})& = & -\frac{x_{1}-y_{1}}{D_{1}-D_{2}}+
\frac{x_{3}-y_{3}}{D_{2}-D_{3}}, \nonumber \\
\frac{1}{2}\frac{d}{dA}\ln(x_{3}-y_{3})& = & -\frac{x_{2}-y_{2}}{D_{2}-D_{3}}+
\frac{x_{1}-y_{1}}{D_{3}-D_{1}}.
\end{eqnarray}
Note that the quantities $D_{i}-D_{j}$ and $x_{i}-y_{i}$ form a
closed system under the evolution equations, independent of other possible 
combinations of these variables.

There remain two more independent evolution equations,
which may be chosen as those for ($x_{i}+y_{i}$). We define
\begin{equation}\label{eq:Xi}
X_{i}=x_{i}-y_{i}, 
\end{equation}
\begin{equation}
\Omega_{i}=x_{i}+y_{i}.
\end{equation}
Then
\begin{eqnarray}
\frac{d\Omega_{i}}{dA}&=&\sum_{j>k}\frac{1}{D_{j}-D_{k}}[\delta_{ij}(\Omega_{i}X_{k}-\Omega_{k}X_{i})
-\delta_{ik}(\Omega_{i}X_{j}-\Omega_{j}X_{i}) \nonumber \\
&-&\epsilon_{ijk}((\Omega_{i}X_{j}-\Omega_{j}X_{i})-(\Omega_{i}X_{k}-\Omega_{k}X_{i}))].
\end{eqnarray}
It follows that 
\begin{equation}
\frac{d}{dA}(x_{i}+y_{i})=0
\end{equation}
if $(x_{j}+y_{j})=0$.  This condition is equivalent to $W_{2i}=W_{3i}$, $i.e.$,
$\mu-\tau$ exchange symmetry.  Thus, the evolution equations preserve the
$\mu-\tau$ symmetry, which was established (approximately) for neutrino
mixing in vacuum.

Another useful property of the evolution equations is to establish matter invariants.
For instance \cite{HaSc,Naumov,Toshev,KrPe},
\begin{equation}\label{xd}
\frac{d}{dA}[X_{1}X_{2}X_{3}\Delta^{2}_{12}\Delta^{2}_{23}\Delta^{2}_{31}]=0,
\end{equation}
where $X_{i}$ is defined in Eq.~(\ref{eq:Xi}) and
\begin{equation}
\Delta_{ij}=D_{i}-D_{j}.
\end{equation}
(Also, $d(J\Delta_{12}\Delta_{23}\Delta_{31})/dA=0$, as mentioned before \cite{CKL,CK10}).   
In addition, there is a simple relation
\begin{equation}\label{sxd}
\frac{1}{2}\frac{d}{dA}\left[\sum_{i>j}(X_{i}-X_{j})\Delta_{ij}\right]=1.
\end{equation}
Eqs. ~(\ref{xd}) and ~(\ref{sxd}) are three-flavor 
generalizations of the two-flavor results \cite{CK10}:  
\begin{equation}
\frac{d}{dA}(xyD^{2})=0,
\end{equation}
\begin{equation}
\frac{d}{dA}[(x+y)D]=-1,
\end{equation}
where $x=V_{11}V_{22}=\cos^{2}\theta$, $y=V_{12}V_{21}=-\sin^{2}\theta$,
$D=m^{2}_{2}-m^{2}_{1}$, in the usual notation.

The vacuum neutrino masses are known to be hierarchical, $\delta_{0}/\Delta_{0} \approx1/32 \ll 1$,
$\delta_{0}=m^{2}_{2}-m^{2}_{1}$, $\Delta_{0} \equiv |m^{2}_{3}-m^{2}_{2}|$.
There are two possibilities, the normal hierarchy ($m^{2}_{3} \gg m^{2}_{1} \approx m^{2}_{2}$),
or the inverted hierarchy ($m^{2}_{3} \ll m^{2}_{1} \approx m^{2}_{2}$).
In matter of constant density, $m_{i}^{2} \rightarrow D_{i}$, which are $A$-dependent.
For the case of normal hierarchy, there are two $A$-values where the levels ``cross",
at the lower resonance, $A=A_{l}$, $[d(D_{1}-D_{2})/dA]_{A_{l}}=0$, and at the higher resonance,
$A=A_{h}$, $[d(D_{2}-D_{3})/dA]_{A_{h}}=0$.  From Eqs.~(\ref{eq:ln}), one finds that rapid variations 
occur only for $A$ to be near $A_{l}$ or $A_{h}$.  Let us denote by
$(A_{0},A_{l},A_{i},A_{h},A_{d})$ the values of $A$ in vacuum $(A_{0}=0)$,
at the lower resonance $(A_{l})$, in the intermediate range $(A_{i})$,
at the higher resonance $(A_{h})$, and in dense medium $(A_{d})$.  Then, the solutions for
$(X,Y)$ are well-approximated \cite{CKL,CK10} by two-flavor resonance solutions.  

For $0<A<A_{i}$,
\begin{eqnarray}\label{low}
\Delta_{21} & = & [p^{2}_{l}A^{2}-2q_{l}\delta_{0}A+\delta_{0}^{2}]^{1/2}, \nonumber \\
X_{1} & = &\frac{1}{2}[p_{l}-(p^{2}_{l}A-q_{l}\delta_{0})/\Delta_{21}], \nonumber \\
X_{2} & = &\frac{1}{2}[p_{l}+(p^{2}_{l}A-q_{l}\delta_{0})/\Delta_{21}], \nonumber \\
X_{3} & \cong & (X_{3})_{0},
\end{eqnarray}
where $\Delta_{ij}=D_{i}-D_{j}$ in matter, $X_{i}=x_{i}-y_{i}$, 
$p_{l}=(X_{1}+X_{2})_{0}$, $q_{l}=(X_{1}-X_{2})_{0}$.  Note that 
$(X_{1})_{0} \cong 2/3$, $(X_{2})_{0} \cong 1/3$, and $X_{1}X_{2}\Delta_{21}^{2}=\mbox{constant}$.

For $A_{i}<A<A_{d}$,
\begin{eqnarray}\label{high}
\Delta_{32}&=&[p^{2}_{h} \bar{A}^{2}-2q_{h}\Delta_{i} \bar{A}+\Delta^{2}_{i}]^{1/2}, \nonumber \\
X_{1} & \cong &(X_{1})_{i}, \nonumber \\
X_{2} &=&\frac{1}{2}[p_{h}-(p^{2}_{h} \bar{A}-q_{h} \Delta_{i})/\Delta_{32}], \nonumber \\
X_{3} &=&\frac{1}{2}[p_{h}+(p^{2}_{h} \bar{A}-q_{h} \Delta_{i})/\Delta_{32}].
\end{eqnarray}
Here, $\bar{A} \equiv A-A_{i}$, and $p_{h}=(X_{2}+X_{3})_{i}$, $q_{h}=(X_{2}-X_{3})_{i}$,
$\Delta_{i}=(\Delta_{32})_{i}$ are taken at $A=A_{i} \gg \delta_{0}$.  
Note that $(X_{1})_{i} \cong 0$,
$(X_{2})_{i} \cong 1$, $(X_{3})_{i} \cong (X_{3})_{0}=|V_{13}|_{0}^{2} \ll 1$.  Also, 
$X_{2}X_{3}\Delta_{32}^{2}$ is an invariant as $A$ varies.  Thus, the product 
$X_{2}X_{3}$ has a resonance behavior near $A\simeq A_{h}$.
Note also that the minimum of $\Delta_{32}$ is at 
$(\Delta_{32})_{min} \simeq 2\sqrt{|V_{13}|_{0}^{2}} \Delta_{0}$.


To obtain $\Delta_{21}$ for $A_{i}<A<A_{d}$ and $\Delta_{32}$ for $0<A<A_{i}$,
one first notes from Eq.~(\ref{eq:di}) that $d\Delta_{21}/dA \simeq X_{2}$ for high $A$.
Thus, a direct integration leads to
\begin{equation}
\Delta_{21}=\delta_{i}+\frac{1}{2}\left[\Delta_{i}+p_{h}\bar{A}-
(p_{h}^{2}\bar{A}^{2}-2q_{h}\Delta_{i}\bar{A}+\Delta_{i}^{2})^{1/2}\right]
\end{equation}
for $A_{i}<A<A_{d}$, where $\delta_{i}=(\Delta_{21})_{i}$.  Similarly, a direct integration of
$d\Delta_{32}/dA \simeq X_{2}$ for low $A$ gives
\begin{equation}
\Delta_{32}=\Delta_{0}+\frac{1}{2}\left[\delta_{0}-p_{l}A
-(p_{l}^{2}A^{2}-2q_{l}\delta_{0}A+\delta_{0}^{2})^{1/2}\right].
\end{equation}
The solutions for $\Delta_{31}$ in both regions of $A$ are obtained from
$\Delta_{31}=\Delta_{32}+\Delta_{21}$.  Note that the solutions 
for $0<A<A_{i}$ and for $A_{i}<A<A_{d}$ should agree
for $A \cong A_{i}$.  This condition leads to $\Delta_{i} \simeq \Delta_{0}-A_{i}$
and $\delta_{i} \simeq A_{i}$.  


For inverted hierarchy, the behaviors of $X_{i}$ near $A_{l}$ are given by the same
Eq.~(\ref{low}).  However, for $A>A_{i}$, there is no longer a resonance.
Instead, all $X_{i}$ change slowly, so that $X_{1} \simeq 0$, $X_{2} \simeq 1$,
$X_{3} \simeq 0$, for $A>A_{i}$.  The solutions for $\bar{\nu}$ are obtained by
$A \rightarrow -A$.  Thus, there is a resonance behavior near $A_{h}$, for the 
inverted hierarchy scenario.  Otherwise all the changes are small.

The accuracy of the approximate formulas in Eqs.~(\ref{low}-\ref{high}) 
can be assessed by numerical integrations of the exact equations,
Eqs.~(\ref{eq:di}) and ~(\ref{eq:ln}).  To do that we write
\begin{equation}\label{w0}
W= \left(\begin{array}{ccc}
  \frac{2(1-\epsilon^{2})}{3}-2\eta & \frac{1-\epsilon^{2}}{3}+2\eta    &  \epsilon^{2} \\
 \frac{1+2\epsilon^{2}-\xi}{6}+\lambda+\eta & 
     \frac{2+\epsilon^{2}-2\xi}{6}-\lambda-\eta & 
     \frac{1-\epsilon^{2}+\xi}{2} \\
 \frac{1+2\epsilon^{2}+\xi}{6}-\lambda+\eta  & 
\frac{2+\epsilon^{2}+2\xi}{6}+\lambda-\eta  & 
    \frac{1-\epsilon^{2}-\xi}{2} \\
                    \end{array}\right),
\end{equation} 
where $(\epsilon, \eta, \lambda, \xi) \ll 1$ in vacuum, and
$W$ reduces to the tribimaximal \cite{tribi} 
matrix when $\epsilon=\eta=\lambda=\xi=0$.

\begin{figure}[ttt]
\caption{The variation of $\Delta_{ij}$ for the normal hierarchy (left) and the inverted hierarchy (right).
Both the numerical (solid lines) and the approximate (dashed lines) solutions are shown.  The approximate analytical
solutions are given by Eqs.(22-25).
We adopt $\lambda=0.02$ and the current upper bound for $|V_{13}|$, $\epsilon \simeq 0.17$.} 
\centerline{\epsfig{file=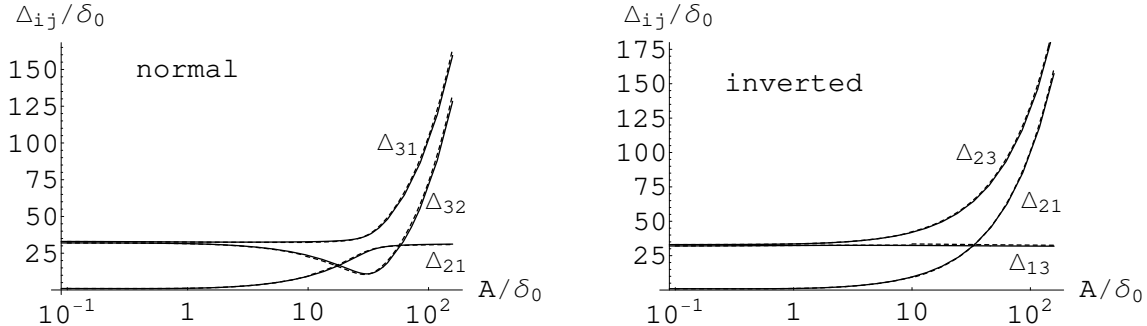,width=16 cm}}
\end{figure} 

It should be emphasized that the parameters 
$(\epsilon, \eta, \lambda, \xi)$ carry quite distinct behaviors as $A$ varies,
as shown in the following.
Eqs.~(\ref{eq:w}) and ~(\ref{w0}) give rise to
\begin{equation}
\xi=W_{23}-W_{33}=(x_{2}+y_{2})-(x_{1}+y_{1}),
\end{equation}
and from $W_{21}-W_{31}$, we have
\begin{equation}
6\lambda=3(x_{3}+y_{3})-2(x_{2}+y_{2})-(x_{1}+y_{1}).
\end{equation}
With the constancy of $x_{j}+y_{j}$, one 
concludes that $\xi\simeq \lambda\approx 0$ as $A$ varies.
In addition, since $W_{11}+W_{12}=1-\epsilon^{2}$, we have
\begin{equation}
\frac{d\epsilon^{2}}{dA}=-\frac{d}{dA}[(x_{2}-y_{2})+(x_{1}-y_{1})],
\end{equation}
and
\begin{eqnarray}
\frac{d\epsilon^{2}}{dA}&=&0,  (\mbox{for low A}) \nonumber \\
\frac{d\epsilon^{2}}{dA} &=& -2(x_{2}-y_{2})(x_{3}-y_{3})/(D_{2}-D_{3}),  (\mbox{for high A})
\end{eqnarray}
Furthermore, one obtains from $W_{12}-W_{11}$ that 
\begin{equation}
\eta =\frac{1}{12}(1-\epsilon^{2})+\frac{1}{4}[(x_{2}-y_{2})-(x_{1}-y_{1})],
\end{equation}
and 
\begin{eqnarray}
\frac{d\eta}{dA}&=&-(x_{2}-y_{2})(x_{1}-y_{1})/(D_{1}-D_{2}), (\mbox{for low A}) \nonumber \\
\frac{d\eta}{dA}&=&\frac{2}{3}(x_{2}-y_{2})(x_{3}-y_{3})/(D_{2}-D_{3}), (\mbox{for high A})
\end{eqnarray}
Thus, $\eta$ and $\epsilon^{2}$ can change considerably as functions of $A$,
but $\lambda \simeq \xi \approx 0$ throughout.

For numerical integrations,
Eqs.~(\ref{eq:w}) and~(\ref{w0}) suggest the following initial values in vacuum:
\begin{eqnarray}\label{eq:initial}
x_{10} & = & \frac{1}{6}(2-3\lambda-2\epsilon^{2}), \hspace{.2in}
y_{10} =  \frac{1}{6}(-2-3\lambda+2\epsilon^{2}), \nonumber \\
x_{20} & = & \frac{1}{6}(1-3\lambda-\epsilon^{2}), \hspace{.2in}
y_{20}  =  \frac{1}{6}(-1-3\lambda+\epsilon^{2}), \nonumber \\
x_{30} & = & \frac{1}{2} (\lambda+\epsilon^{2}), \hspace{.2in}
y_{30}  =  \frac{1}{2}(\lambda-\epsilon^{2}),
\end{eqnarray}
where $\xi=\eta=0$ is chosen and the terms in $\mathcal{O}(\lambda \epsilon^{2})$
are ignored.  We shall choose the initial values 
$\epsilon=0.17$ and $\lambda =0.02$, which correspond to the experimental bounds
$|V_{13}|^{2} \leq 0.03$ \cite{data} and an assumed CP violation
phase $\cos \varphi=1/4$, respectively.
The numerical solutions for the $(x,y)$ parameters, the squared elements of
the mixing matrix, and $J$ in matter follow directly and are shown
in Figs. 2-5 in Ref. \cite{CK10}.
Our choice of $\lambda \neq 0$ signifies a small $\mu-\tau$ symmetry breaking,
the solutions verify that $(x_{i}+y_{i})$ remain negligible for all A values.
In addition, we show in Fig. 1 both the numerical and the approximate solutions for $\Delta_{ij}$ in matter.
Note that the hierarchical relation among the $\Delta_{ij}$'s varies in matter and plays
an important role in the oscillatory factor $\sin^{2}\Phi_{ij}$ of the probability functions.
It is seen that 
$\Delta_{21}/\Delta_{31} \simeq \Delta_{21}/\Delta_{32}\sim 1/32 \ll 1$ (normal hierarchy)
and $\Delta_{21}/\Delta_{23} \simeq \Delta_{21}/\Delta_{13} \sim1/32 \ll 1$ 
(inverted hierarchy) for $0<A \lesssim A_{i}$. 
While in $A_{i} < A \lesssim A_{d}$,
the $\Delta_{ij}$'s are less hierarchical: 
$\Delta_{21}/\Delta_{31} \sim \Delta_{21}/\Delta_{32} \gtrsim 1/5$ (normal)
and $\Delta_{13}/\Delta_{21} \sim \Delta_{13}/\Delta_{23} \gtrsim 1/5$ (inverted).



\begin{table*}[ttt]
  \centering
	\begin{center}
 \begin{tabular}{ccccc}  
 
 \hline

     $F^{\alpha \beta}_{ij}$ & complete & with $x_{i}+y_{i}\simeq 0$ & & $A_{i}<A<A_{d}$    \\ \hline
    $F^{e \mu}_{21}$ & $-x_{1}x_{2}-x_{1}x_{3}+x_{1}y_{2}+y_{1}y_{3}$  & 
    $-2x_{1}x_{2}$ & &   $\ll 1$     \\
    $F^{e \mu}_{31}$ & $x_{1}x_{2}+x_{3}y_{1}-y_{1}y_{2}-y_{1}y_{3}$  & 
    $-2x_{1}x_{3}$ & &  $\ll 1$     \\
    $F^{e \mu}_{32}$ & $-x_{1}x_{2}-x_{2}x_{3}+x_{2}y_{3}+y_{1}y_{2}$  & 
    $-2x_{2}x_{3}$ &  & $-2x_{2}x_{3}$    \\    
      $F^{e \tau}_{21}$ & $+x_{1}x_{3}+x_{2}y_{1}-y_{1}y_{2}-y_{1}y_{3}$  & 
    $-2x_{1}x_{2}$ & & $\ll 1$     \\
    $F^{e \tau}_{31}$ & $-x_{1}x_{2}-x_{1}x_{3}+x_{1}y_{3}+y_{1}y_{2}$  & 
    $-2x_{1}x_{3}$ & & $\ll 1$     \\
    $F^{e \tau}_{32}$ & $-x_{1}x_{2}+x_{3}y_{2}-y_{1}y_{2}-y_{2}y_{3}$  & 
    $-2x_{2}x_{3}$ &  & $-2x_{2}x_{3}$    \\                 
     $F^{\mu \tau}_{21}$ & $-x_{1}x_{3}-x_{2}x_{3}+x_{3}y_{3}+y_{1}y_{2}$  & 
    $-(x_{3}/2)+x_{1}x_{2}$ & & $-x_{3}/2$     \\
    $F^{\mu \tau}_{31}$ & $x_{1}x_{3}+x_{2}y_{2}-y_{1}y_{2}-y_{2}y_{3}$  & 
    $-(x_{2}/2)+x_{1}x_{3}$ & & $-x_{2}/2$     \\
    $F^{\mu \tau}_{32}$ & $-x_{1}x_{2}-x_{1}x_{3}+x_{1}y_{1}+y_{2}y_{3}$  & 
    $-(x_{1}/2)+x_{2}x_{3}$ & & $x_{2}x_{3}$    \\  
   \hline
  \end{tabular}
    \caption{The complete and the approximate forms for
     the functions $F^{\alpha \beta}_{ij}$ in all channels under the normal hierarchy.  Note that 
     the approximation for $F^{\alpha \beta}_{ij}$ in $0<A \lesssim A_{i}$ is 
     the same as that with $x_{i}+y_{i}\simeq 0$ and is omitted.  Note also that
     $F^{\alpha\beta}_{ij}=F^{\beta\alpha}_{ij}$.}  
  \end{center}
 \end{table*}




\section{The probability functions}

The neutrino transition probability in matter is given by \cite{data}
\begin{eqnarray}
P(\nu_{\alpha} \rightarrow \nu_{\beta})&=&\delta_{\alpha \beta}-
4\sum_{j>i}\mbox{Re}(\Pi^{\alpha \beta}_{ij})
\sin^{2}\Phi_{ij} \nonumber \\
 &+& 2\sum_{j>i}\mbox{Im}(\Pi^{\alpha \beta}_{ij})
\sin2\Phi_{ij},
\end{eqnarray}
where $\Pi^{\alpha \beta}_{ij} \equiv V_{\alpha i}V_{\beta j}V^{*}_{\alpha j}V^{*}_{\beta i}$
(Eq.~(\ref{Pi})), and
\begin{equation}
\Phi_{ij} \equiv \Delta_{ij} L/4E,
\end{equation}
with $L=$ baseline length.  We can rewrite the probability functions in terms of the physical observables
$(x,y)$.  Let us write, for $\alpha \neq \beta$,
\begin{eqnarray}\label{eq:ab}
P(\nu_{\alpha} \rightarrow \nu_{\beta})&=&-4[F^{\alpha \beta}_{21} \sin^{2}\Phi_{21}+ 
F^{\alpha \beta}_{31} \sin^{2}\Phi_{31} 
+F^{\alpha \beta}_{32} \sin^{2}\Phi_{32}] \nonumber \\
   &-& 8J\sin \Phi_{21}\sin \Phi_{31}\sin \Phi_{32}.
  \end{eqnarray}
Using Eq.~(\ref{Pi2}), $F^{\alpha \beta}_{ij}$ can all be expressed as
quadratic forms in $(x,y)$.  They are listed in Table I.
The functions $F^{\alpha \beta}_{ij}$ can be further simplified by
using the approximate $\mu-\tau$ symmetry, $x_{i}+y_{i} \simeq 0$, for all $A$.
In addition, with normal hierarchy, $x_{1} \ll 1$ for $A_{i}<A<A_{d}$.
These approximate results are also listed in Table I.
Despite the fact that $x_{3} \ll 1$ for $0 \lesssim A \lesssim A_{i}$,
terms containing $x_{3} \simeq |V_{13}|^{2}/2$ are kept so that the physical
potential can be explored.  Finally, it is noteworthy that
the term $x_{2}x_{3}$, according to Eq.~(\ref{high}), has a resonance
behavior near $A\simeq A_{h}$.  This is a distinctive feature that can be exploited by
proper choices of parameters in an experiment.

For $\alpha=\beta$, we write
\begin{equation}
P(\nu_{\alpha} \rightarrow \nu_{\alpha})
=1-\sum_{\alpha \neq \beta} P(\nu_{\alpha} \rightarrow \nu_{\beta})
=1+4\sum_{j>i}F^{\alpha \alpha}_{ij} \sin^{2}\Phi_{ij}.
\end{equation}
We list $F^{\alpha \alpha}_{ij}$ in Table II.


\begin{table*}[ttt]
  \centering
	\begin{center}
 \begin{tabular}{ccccc}  
 
 \hline

     $F^{\alpha \alpha}_{ij}$ & complete & with $x_{i}+y_{i}\simeq 0$ & & $A_{i}<A<A_{d}$    \\ \hline
    $F^{ee}_{21}$ & $-x_{1}x_{2}+x_{1}y_{2}+x_{2}y_{1}-y_{1}y_{2}$  & 
    $-4x_{1}x_{2}$  & &  $\ll 1$     \\
    $F^{ee}_{31}$ & $-x_{1}x_{3}+x_{1}y_{3}+x_{3}y_{1}-y_{1}y_{3}$  & 
    $-4x_{1}x_{3}$ &  & $\ll 1$     \\
    $F^{ee}_{32}$ & $-x_{2}x_{3}+x_{2}y_{3}+x_{3}y_{2}-y_{2}y_{3}$  & 
    $-4x_{2}x_{3}$  &  & $-4x_{2}x_{3}$    \\                  
    $F^{\mu \mu}_{21}$ & $-x_{1}x_{3}+x_{3}y_{3}+x_{1}y_{2}-y_{2}y_{3}$  & 
    $-x_{1}x_{2}-(x_{3}/2)$  &  & $-x_{3}/2$     \\
    $F^{\mu \mu}_{31}$ & $-x_{2}x_{3}+x_{3}y_{1}+x_{2}y_{2}-y_{1}y_{2}$  & 
    $-x_{1}x_{3}-(x_{2}/2)$  & & $-x_{2}/2$     \\
    $F^{\mu \mu}_{32}$ & $-x_{1}x_{2}+x_{1}y_{1}+x_{2}y_{3}-y_{1}y_{3}$  & 
    $-x_{2}x_{3}-(x_{1}/2)$  & & $-x_{2}x_{3}$    \\  
      $F^{\tau \tau}_{21}$ & $-x_{2}x_{3}+x_{2}y_{1}+x_{3}y_{3}-y_{1}y_{3}$  & 
    $-x_{1}x_{2}-(x_{3}/2)$  & & $-x_{3}/2$     \\
    $F^{\tau \tau}_{31}$ & $-x_{1}x_{2}+x_{2}y_{2}+x_{1}y_{3}-y_{2}y_{3}$  & 
    $-x_{1}x_{3}-(x_{2}/2)$  & & $-x_{2}/2$     \\
    $F^{\tau \tau}_{32}$ & $-x_{1}x_{3}+x_{3}y_{2}+x_{1}y_{1}-y_{1}y_{2}$  & 
    $-x_{2}x_{3}-(x_{1}/2)$ & & $-x_{2}x_{3}$ \\
      \hline
  \end{tabular}
    \caption{The complete and approximate forms for the 
    functions $F^{\alpha \alpha}_{ij}$ in all channels under the normal hierarchy.}
  \end{center}
 \end{table*}


Our results may be compared to formulas in terms of the ``standard parametrization" \cite{data},  
given, $e.g.$, in Kimura $\emph{et al.}$ \cite{Ki}.  
The relations between $(x,y)$ and the ``standard parametrization" are given by
\begin{eqnarray}\label{eq:KK}
J&=&K\sin\varphi \nonumber \\
K&=&s_{12}c_{12}s_{13}c^{2}_{13}s_{23}c_{23} \nonumber \\
x_{1}&=&c^{2}_{12}c^{2}_{13}c^{2}_{23}-K\cos\varphi \nonumber \\
x_{2}&=&s^{2}_{12}c^{2}_{13}s^{2}_{23}-K\cos\varphi \nonumber \\
x_{3}&=&s^{2}_{12}s^{2}_{13}c^{2}_{23}+c^{2}_{12}s^{2}_{13}s^{2}_{23}+
      \frac{1+s^{2}_{13}}{1-s^{2}_{13}}K\cos\varphi \nonumber \\
y_{1}&=&-c^{2}_{12}c^{2}_{13}s^{2}_{23}-K\cos\varphi \nonumber \\
y_{2}&=&-s^{2}_{12}c^{2}_{13}c^{2}_{23}-K\cos\varphi \nonumber \\
y_{3}&=&-s^{2}_{12}s^{2}_{13}s^{2}_{23}-c^{2}_{12}s^{2}_{13}c^{2}_{23}+
      \frac{1+s^{2}_{13}}{1-s^{2}_{13}}K\cos\varphi
\end{eqnarray}
where $s_{ij}\equiv \sin\theta_{ij}$, $c_{ij}\equiv\cos\theta_{ij}$, 
and $\varphi$ is the Dirac CP phase. 
It can be shown that the functions
$F^{\alpha\beta}_{ij}$ here in terms of $(x,y)$ are simply
$\mbox{Re}J^{ij}_{\alpha\beta}$ in Eqs.(15-23) of Ref. \cite{Ki}, and the resultant
probability functions are identical.
Eq~(\ref{eq:KK}) also offers some insight on the $A$-independence of the approximate
$\mu-\tau$ symmetry.  It is seen that the conditions $x_{i}+y_{i}=0$ are fulfilled if
1) $c^{2}_{23}=s^{2}_{23}$, and 2) $s_{12}c_{12}s_{13}s_{23}c_{23}\cos\varphi =0$.
The behaviors of $s_{ij}$ were given in Fig. 6 of Ref. \cite{CK10}.  
While $s^{2}_{23}\cong 1/2$ is almost independent of $A$, $s_{13}\cong 0$
for low $A$, and $c_{12} \cong 0$ for high $A$.  They combine to validate conditions
1) and 2), for all $A$ values.    
The other possibility is that $\cos\varphi =0$.  Here, $\varphi$ itself is
largely $A$-independent because of the matter invariant 
$\sin\varphi\sin 2\theta_{23}$ \cite{Toshev}.

Exact $\mu-\tau$ symmetry was studied earlier by Harrison and Scott \cite{23-sym}.
Their formulation uses the mixing matrix $V$ (with specific choice of phases),
while our results are in terms of rephasing invariant (and observable) variables,
making it possible to calculate transition probabilities directly.
In addition, by comparing with the exact formulas in Table I, one can
quickly compute corrections to the presumed exact symmetry.

\begin{figure}[ttt]
\caption{For $L=2540$ km, the resonant location of $8x_{2}x_{3}$ and the peak of the
oscillatory factor $\sin^{2}\Phi_{32}$ do not coincide, and the resultant
probability $P\simeq 8x_{2}x_{3}\sin^{2}\Phi_{32}$ is suppressed.
Note that the probability $P\simeq 8x_{2}x_{3}\sin^{2}\Phi_{32}$ is shown here
for a check of the qualitative property at high energy. The large, fast oscillating
probability near the low energy is not seen here because the $x_{1}x_{2}$ and $x_{1}x_{3}$
terms are ignored in $P\simeq 8x_{2}x_{3}\sin^{2}\Phi_{32}$.} 
\centerline{\epsfig{file=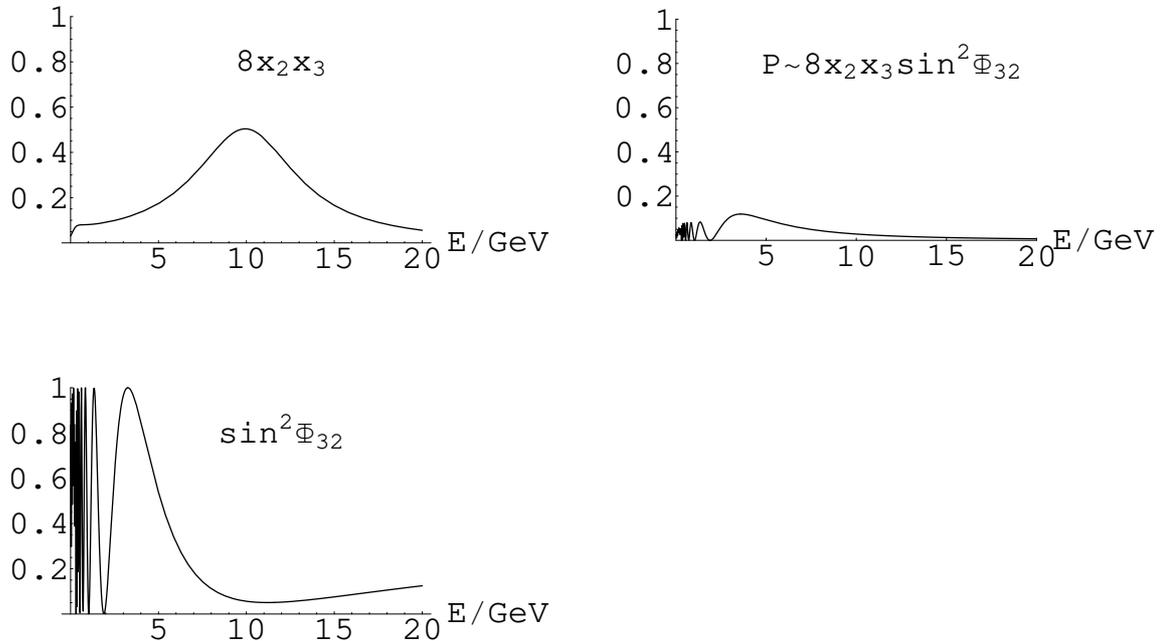,width=16 cm}}
\end{figure} 


\section{Applications to the long baseline experiments}

The unique features of the $(x,y)$ parametrization can be used to facilitate,
$e.g.$, the analyses of the LBL experiments.  As an example, let us consider the
probability $P(\nu_{e}\rightarrow \nu_{\mu})$ explicitly.
According to Table I, with the approximation $x_{i}+y_{i}=0$, 
\begin{eqnarray}
P(\nu_{e} \rightarrow \nu_{\mu})&=& 8[x_{1}x_{2} \sin^{2}\Phi_{21}+x_{1}x_{3} \sin^{2}\Phi_{31}+
x_{2}x_{3} \sin^{2}\Phi_{32}] \nonumber \\
&-&8J\sin\Phi_{21}\sin\Phi_{31}\sin\Phi_{32},
\end{eqnarray}
with $J=\pm \sqrt{2x_{1}x_{2}x_{3}}$.  Using the solutions in Eqs.~(\ref{low},\ref{high}),
it is straightforward to infer the behaviors of $P(\nu_{e}\rightarrow \nu_{\mu})$.
In the following, let us focus on the region of high $A$ values $(A_{i}<A<A_{d})$.
Here, $x_{1} \ll 1$ so that (excluding the case $\Phi_{32} \ll1$)
\begin{equation}\label{pemu}
P(\nu_{e} \rightarrow \nu_{\mu}) \simeq 8x_{2}x_{3} \sin^{2}\Phi_{32}.
\end{equation}
It is useful to examine the qualitative properties of $x_{2}x_{3}$ and $\sin^{2}\Phi_{32}$ separately. 
If the mass hierarchy is normal, 
the solutions in Eq.~(\ref{high}) suggest a higher resonance for $x_{2}x_{3}$ at
$A_{h} \simeq (q_{h}/p_{h}^{2})\Delta_{0} \simeq \Delta_{0}$, where
$\Delta_{0}/\delta_{0} \approx 32$.  
With $A/\delta_{0} \simeq [\rho/(\mbox{g/cm}^{3})][(E/\mbox{GeV})]$,
$\delta_{0} \approx 7.6 \times 10^{-5}$ $\mbox{eV}^{2}$, and $\rho \approx 3.0$ $\mbox{g/cm}^{3}$,
the location of resonance $A_{h}$ corresponds to an energy $E_{h} \sim 10$ GeV,
which is independent of the baseline length.  
Eq.~(\ref{pemu}) shows that, in the high $A$ region, 
$P(\nu_{e}\rightarrow \nu_{\mu})$ $(\cong P(\nu_{e}\rightarrow \nu_{\tau}))$
is two-flavor like.  However, it does not mean that the three-flavor
problem is reduced to a single two-flavor problem.  This is because
the probability $P(\nu_{\mu}\rightarrow \nu_{\tau})$, according to Table I,
would have contributions from all the $\Phi_{ij}$'s.


As an illustration, we show $8x_{2}x_{3}$, $\sin^{2}\Phi_{23}$, and 
$P(\nu_{e}\rightarrow \nu_{\mu})\simeq 8x_{2}x_{3}\sin^{2}\Phi_{23}$ 
as functions of $E$ in Fig. 2, with $L=2540$ km.
It is seen that a resonance for $8x_{2}x_{3}$ occurs near $E \cong 10$ GeV as expected.
However, the smallness of $\sin^{2}\Phi_{32}$ near $E \cong 10$ GeV suppresses the probability even if
$8x_{2}x_{3}$ is at a resonance.
On the other hand, the probability at the first peak of $\sin^{2}\Phi_{32}$ (near $E \cong 3.5$ GeV)
also gets suppressed by the smallness of $8x_{2}x_{3}$.
As a result, a significant flavor transition only occurs when $L$ is adjusted
so that the peak of $\sin^{2}\Phi_{23}$ 
is located near the resonance of $8x_{2}x_{3}$.


\begin{figure}[ttt]
\caption{The probability $P(\nu_{e} \rightarrow \nu_{\mu})$ as a function of $E$
under the normal hierarchy (left) and the inverted hierarchy (right).  It is seen that
$P(L_{1}=7500$ km) $\gg P(L_{2}=750$ km) near the first peak under the normal hierarchy,
while $P(L_{1}=7500$ km) $\approx P(L_{2}=750$ km) $\ll 1$ for the inverted hierarchy.
Note that the chosen baseline $L_{1}=7500$ km leads to the first peak at $E \cong 8$ GeV.} 
\centerline{\epsfig{file=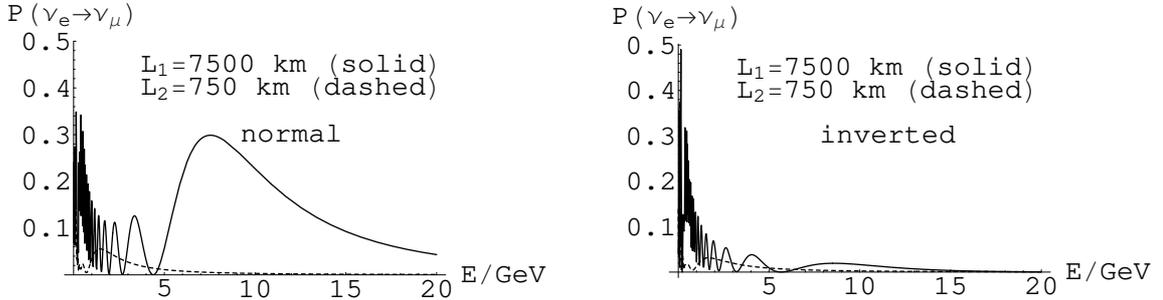,width=16 cm}}
\end{figure} 

The first maximum of $\sin^{2}\Phi_{32}$ occurs if $L/E$ is properly chosen:
\begin{equation}
\Phi_{32}=\Delta_{32}(\frac{L}{4E}) \simeq 9.65 
\times 10^{-5}(\frac{\Delta_{32}}{\delta_{0}})[\frac{(L/\mbox{km})}{(E/\mbox{GeV})}]=\frac{\pi}{2}.
\end{equation}
For the first maximum to coincide with the resonance of $x_{2}x_{3}$,
the value of $\Delta_{32}$ is taken at $A_{h}$: 
$\Delta_{32}/\delta_{0} \simeq 2\sqrt{|V_{13}|^{2}_{0}}\Delta_{0}/\delta_{0} \cong 11$. 
It leads to $(L/\mbox{km})/(E/\mbox{GeV})\sim 10^{3}$  
using the current upper bound $|V_{13}|^{2}_{0} \sim 0.03$. 
One concludes that if the mass hierarchy is normal,
an extra long baseline ($L \sim 10^{4}$ km) can lead to a greatly enhanced probability 
for the neutrino beam near $E \sim 10$ GeV, at which energy both $8x_{2}x_{3}$ and
$\sin^{2}\Phi_{32}$ reach the maximal values.  The probability will be suppressed
when $L$ starts to vary and $\sin^{2}\Phi_{32}$ moves away from the maximum.
Note that for the maxima of $x_{2}x_{3}$ and $\sin^{2}\Phi_{32}$ to coincide near $E \sim 10$ GeV,
the baseline $L$ and the undetermined $|V_{13}|^{2}_{0}$ are related by 
$(L/km)(|V_{13}|_{0}) \sim 2.54 \times 10^{3}$.

On the other hand,
since $8x_{2}x_{3}$ does not go through the higher resonance under the inverted hierarchy,
the probability is in general suppressed even if $\sin^{2}\Phi_{32}$ reaches its maximum.
One further concludes that under the inverted hierarchy,
the transition probability remains small and is insensitive to 
variation of the baseline length $L$.

Thus, if the mass hierarchy is normal,
one would expect to observe sizable probability difference at high energy for experiments
involving two baselines with sizable difference in length.  On the other hand, 
the probability would be small and nearly independent
of the baseline at high energy if the mass hierarchy is inverted.  
We show in Fig. 3 the probability function under both hierarchies for 
two arbitrarily chosen baselines. 
Note that the peak locations and the peak values vary as $L$.
It is seen that for the normal hierarchy, 
$P(L_{1}=7500$ km) $\gg P(L_{2}=750$ km) near the first peak is expected,
while $P(L_{1}=7500$ km) $\approx P(L_{2}=750$ km) $\ll 1$ if the mass hierarchy is inverted.
This result may provide useful hints to the determination of the mass hierarchy.
Note that the probabilities can be deduced
if the details of the experiments are considered. If the neutrino energy can be 
reconstructed accurately from the secondary particles involved in an experiment,
the observed spectrum will tell how the magnitude of the
transition probability plays a role.  On the other hand, 
if reliable measurement of the energy spectrum is not
available, a collection of the event rates should also be useful in
comparing the probabilities.


\begin{figure}[ttt]
\caption{With the baseline $L=5000$ km, 
$P(\nu_{e} \rightarrow \nu_{\mu})$ and $P(\bar{\nu}_{e} \rightarrow \bar{\nu}_{\mu})$
are compared under both the normal and the inverted hierarchies.
It is seen that near the first peak, 
$P(\nu_{e} \rightarrow \nu_{\mu})/P(\bar{\nu}_{e} \rightarrow \bar{\nu}_{\mu}) \gg 1$
if the hierarchy is normal, 
while $P(\nu_{e} \rightarrow \nu_{\mu})/P(\bar{\nu}_{e} \rightarrow \bar{\nu}_{\mu}) \ll 1$
if the hierarchy is inverted.  
Note that $\lambda=0.02$ and the current upper bound for $|V_{13}|$, $\epsilon=0.17$ have been used.} 
\centerline{\epsfig{file=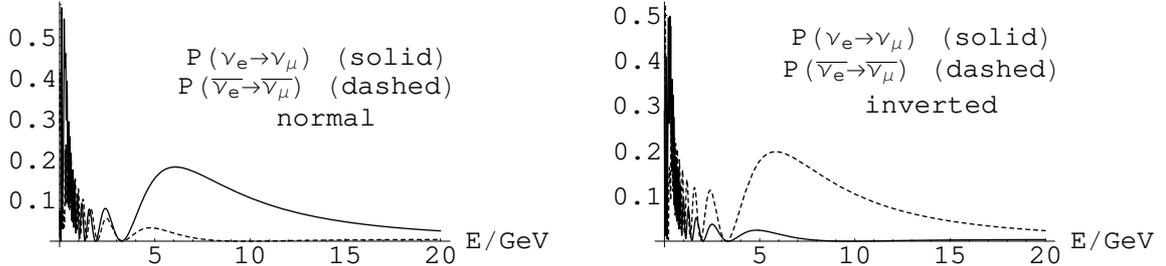,width=16 cm}}
\end{figure} 

Another possible application is to look for both $P(\nu_{e} \rightarrow \nu_{\mu})$
and $P(\bar{\nu}_{e} \rightarrow \bar{\nu}_{\mu})$ for a single, but very long baseline.
Since the $\bar{\nu}$'s only go through the higher resonance under the inverted hierarchy,
one would expect to observe in the vicinity of the peak either
$P(\nu_{e} \rightarrow \nu_{\mu})/P(\bar{\nu}_{e} \rightarrow \bar{\nu}_{\mu}) \gg 1$
if the hierarchy is normal, 
or $P(\nu_{e} \rightarrow \nu_{\mu})/P(\bar{\nu}_{e} \rightarrow \bar{\nu}_{\mu}) \ll 1$
if the hierarchy is inverted. We show an example in Fig. 4.
Note that although the peak value of the probability varies with the baseline length,
the relative and qualitative features of the above observation remain valid for a chosen baseline.


\section{Conclusions}     
 
Neutrino transition probabilities are usually given in terms of the simple expression
$(V_{\alpha i}V_{\beta j}V^{*}_{\alpha j}V^{*}_{\beta i})$, although the individual 
$V_{\alpha i}$'s are not directly observable.    When one rewrites them using physical observables,
such as those in the ``standard parametrization", the resulting formulas are often very
complicated.  It is thus not easy to obtain general properties of these probabilities in
experimental situations.  In this paper we express the probabilities
as functions of rephasing invariant parameters.  
In addition, we incorporate the $\mu-\tau$ symmetry, valid (approximately)
for any value of the induced neutrino mass ($A$).  The resulting formulas are very simple, 
and are listed in Tables I and II.  They offer a quick quantitative assessment for any
physical process at arbitrary $A$ values.  As an illustration, we analyzed the probability
$P(\nu_{e} \rightarrow \nu_{\mu})$, with emphasis on its dependence on $E$, $L$,
and $\sqrt{|V_{13}|^{2}_{0}}$.  By changing the value of $E$ and $L$ in various LBL experiments,
one can hope not only to test the theory used to establish 
$P(\nu_{\alpha} \rightarrow \nu_{\beta})$, but also to help in the 
efforts to determine the unknown parameter $|V_{13}|^{2}_{0}$.

$Acknowledgments$ SHC is supported by the National 
Science Council of Taiwan, Grant No. NSC 98-2112-M-182-001-MY2.

\end{document}